\documentclass[aps,prl,twocolumn,nopacs,lettersize,superscriptaddress]{revtex4-1}
\usepackage{hyperref}
\usepackage{graphicx}
\usepackage{array}
\usepackage{rotating}
\usepackage{multirow}
\usepackage{placeins}
\usepackage{color}
\usepackage[mediumspace,mediumqspace,Grey,squaren]{SIunits}
\usepackage{notes2bib}

\begin{document}

\title{Deterministic creation, pinning, and manipulation of quantized vortices in a Bose-Einstein condensate}


\author{E.~C.~Samson}
\altaffiliation[Current address: ]{Physics Division, Los Alamos National Laboratory, Los Alamos, NM 87545, USA}
\affiliation{College of Optical Sciences, University of Arizona, Tucson, AZ 85721 USA}
\author{K.~E.~Wilson}
\altaffiliation[Current address: ]{School of Engineering and Physical Sciences, David Brewster Building, Heriot-Watt University, Edinburgh, Scotland EH14 4AS, United Kingdom}
\affiliation{College of Optical Sciences, University of Arizona, Tucson, AZ 85721 USA}
\author{Z.~L.~Newman}
\affiliation{College of Optical Sciences, University of Arizona, Tucson, AZ 85721 USA}
\author{B.~P.~Anderson}
\email[Email: ]{bpa@optics.arizona.edu}
\affiliation{College of Optical Sciences, University of Arizona, Tucson, AZ 85721 USA}

\date{\today}

\begin{abstract}

We experimentally and numerically demonstrate deterministic creation and manipulation of a pair of oppositely charged singly quantized vortices in a highly oblate Bose-Einstein condensate (BEC).  Two identical blue-detuned, focused Gaussian laser beams that pierce the BEC serve as repulsive obstacles for the superfluid atomic gas; by controlling the positions of the beams within the plane of the BEC, superfluid flow is deterministically established around each beam such that two vortices of opposite circulation are generated by the motion of the beams, with each vortex pinned to the \emph{in situ} position of a laser beam.   We study the vortex creation process, and show that the vortices can be moved about within the BEC by translating the positions of the laser beams.   This technique can serve as a building block in future experimental techniques to create, on-demand, deterministic arrangements of few or many vortices within a BEC for precise studies of vortex dynamics and vortex interactions.
\end{abstract}


\maketitle

\section{I. Introduction}

Quantized vortices are central features of macroscopic quantum coherent phenomena \cite{Don1991.Vortices.book,Fet2001.JPCM13.R135,Ann2004.Superconductivity.book, Fet2010.JLTP161.445}, and serve as robust indicators of the dynamics of quantum fluids.  Through the development of progressively better experimental techniques for controlling and observing the motion of vortices in these systems, new possibilities open up to better understand the physics of vortices in a quantum fluid.  In turn, new techniques for controlling the states and properties of quantum fluids may soon follow.   For example, in type-II superconductors immersed in a strong magnetic field, the motion of magnetic flux vortices is largely responsible for energy dissipation and limiting the temperatures at which superconductivity in these materials occurs \cite{Ann2004.Superconductivity.book}; by pinning or otherwise arresting the motion of vortices, superconductivity may persist at higher temperatures and magnetic fields than if the vortices were free to move.  New research on the parameters that influence vortex motion may thus help extend the temperature and magnetic field ranges of such superconductors \cite{Cor2013.NatCom4.1437}.  In superfluid helium, vortex-vortex interactions and Kelvin waves on vortex lines also dissipate energy; new methods to measure and better understand the dynamics of vortices in a superfluid might shed new information on quantum turbulence energy dissipation mechanisms in these systems \cite{Fon2014.PNAS111.4707}.  Similarly, in two-dimensional (2D) quantum turbulence in highly oblate Bose-Einstein condensates (BECs) \cite{Nee2013.PRL111.235301}, vortex motion is a key indicator of the hydrodynamic state of the BEC; studies of vortices and their motion in these systems may aid in the development of a deeper understanding of 2D quantum turbulence \cite{Ree2014.PRA89.053631,Bil2014.PRL112.145301}.   To better understand these and other quantum fluid phenomena, it is essential to construct a detailed understanding of vortices, their dynamics, and their interactions under a wide variety of conditions and in different systems.

Dilute-gas BECs are well suited to experimental, theoretical, and numerical studies of vortices, and to direct quantitative comparison between experimental and numerical results that enable rapid development of our understanding of vortices and superfluidity in BECs.  With the publication of over 100 articles to date covering experimental work with vortices since 1999, the experimental techniques available for creating and studying vortices in BECs are by now numerous and widely varied \cite{And2010.JLTP161.574,And.BEC.vortex}.    In most of these studies, the specific positions of each vortex within the BEC are not of primary concern, and vortex creation is not deterministic at the single-vortex level.   Nevertheless, deterministic vortex creation is possible, and some experiments have indeed focused on deterministic creation mechanisms and even pinning of a singly quantized or multiply quantized vortex or establishment of a persistent current within a BEC \cite{Mat1999.PRL83.2498,Lea2002.PRL89.190403,Kum2006.PRA73.063605,And2006.PRL97.170406,Ryu2007.PRL99.260401,Wri2009.PRL102.030405,Ram2011.PRL106.130401,Cho2012.NJP14.053013,Mou2012.PRA86.013629,Bea2013.PRL110.025301,Wri2013.PRL110.025302,Ryu2014.NJP16.013046,Ray2014.Nat505.657}.   Vortices in BECs are also large enough to be detected optically, either with \emph{in situ} techniques involving multi-component BECs \cite{Mat1999.PRL83.2498} or single-component BECs \cite{Wil2015.PRA91.023621}, or most commonly with a now-standard method of imaging the BEC after a short period of ballistic expansion \cite{Mad2000.PRL84.806}.    Various techniques enable the dynamics of vortices to be measured, either in real-time \cite{Fre2010.Sci329.1182,Mid2011.PRA84.011605} or after deterministic vortex-dipole nucleation \cite{Nee2010.PRL104.160401} for single-component BECs, or with \emph{in situ} imaging for multi-component BECs \cite{Mat1999.PRL83.2498}.  Vortex dynamics and pinning are also strongly affected by the presence of time-dependent potential barriers \cite{Law2014.PRA89.053606}, and even arrays of vortices can be pinned in a rotating frame by laser beams \cite{Tun2006.PRL97.240402}.

However, despite such a proliferation of experiments with vortices in BECs, there are still no published experimental techniques that have demonstrated the flexibility of constructing arbitrary vortex configurations in a BEC, such that the location, vorticity, and circulation direction for each individual vortex can be presicribed deterministically.    Such a technique would be an important component of precision studies of vortex dynamics and interactions, and in manipulating and studying states of a quantum fluid.   Especially under conditions where existing single-shot imaging techniques are used, the ability to specify vortex distributions on-demand would open up new possibilities for studying vortex dynamics and interactions.    To address the challenge of developing such a technique, we present here a method of deterministically generating and manipulating two singly quantized vortices of opposite circulation in a highly oblate BEC, showing results of an experimental study and numerical simulations.   We expect that this technique is scalable well beyond two vortices, and in principle may be used to create designer distributions of many vortices; this possibility is numerically explored in a companion article \cite{Ger2015.ARXIV15xx.xxxx}.   

Below, we first present the conceptual foundation for our technique, followed by a description of our experimental setup and the beam parameters used in our study.  To illustrate the vortex creation process described above, we then show results of numerical simulations of the 2D Gross-Pitaevskii equation for conditions closely approximating those of our experiment.  Finally, we discuss various aspects of the manipulation of vortices within the BEC.


\section{II. Concept}

Our technique generally proceeds as follows.  We begin with two blue-detuned laser beams that pierce a stationary BEC at positions that can be controlled electronically.   The beams are initially co-located, then simultaneously and at a constant speed move together across the BEC, say in the $x$-direction, while also moving slightly apart in the $y$-direction before coming to rest at positions for which the beams are fully separated.  The laser beams serve as repulsive obstacles for the atomic superfluid, pushing fluid out of their way as fluid simultaneously fills the space vacated by the beams during their motion.  The beam motion results in macroscopic flow within the BEC, even for conditions where the speed of the laser beams is far too slow for vortex dipole nucleation and shedding into the bulk superfluid.

As the beams move through the BEC, they eventually separate and become two moving obstacles, allowing the flow that was occupying the space behind the combined beams to merge with the portion of the fluid that was being pushed out of the way ahead of the beams.   For a suitable range of beam trajectory parameters, the now-continuous fluid flow around each separate beam corresponds to a net winding of the BEC's quantum phase of $\pm2\pi$.   In other words, at the moment of beam separation, two opposite-circulation regions of persistent superfluid flow are trapped around each laser beam.  Equivalently, this process can be seen as the simultaneous nucleation and pinning of two oppositely charged vortices, with one vortex pinned to each beam.    Subsequent motion of the laser beams can transport vortices to other locations within the BEC, reminiscent of the use of lasers as optical tweezers for material particles, and allowing for deterministic vortex creation and location.    To emphasize (i) the need to use laser beam pairs in our new method, (ii) the two different linear trajectories that the beams take when moving through the BEC (described in detail below), and (iii) the vortex pinning and manipulation capabilities provided by the beams, we henceforth refer to this dual-vortex creation and manipulation method as the ``chopsticks'' technique.


\section{III.  Experiment}

We create BECs of $^{87}$Rb in the $5\,^2S_{1/2}$ $|F=1,\,m_F=-1\rangle$ state in a highly oblate harmonic trap.  The trapping potential is a combination of a magnetic time-averaged orbiting potential (TOP) magnetic trap \cite{Pet1995.PRL74.3352} with an axis of symmetry in the vertical ($z$) direction, and a red-detuned 1090-nm laser beam that propagates horizontally along the $x$ direction and is tightly focused and provides tight confinement in the $z$ direction.   The harmonic trapping frequencies of this hybrid trap are $(\omega_r/2\pi,\omega_z/2\pi)=(8, 90)$ Hz in the radial and vertical directions, respectively.   With no laser beams present other than the 1090-nm laser beam used for the hybrid trap, our BECs have Thomas-Fermi radii of $R_r \sim$ \unit{50}{\micro\meter} in the radial $(x,y)$ plane, corresponding to condensates of about $N_c \sim 2 \times 10^6$  atoms and a chemical potential of $\mu_0\sim8\,\hbar\omega_z$.  Further details of our BEC creation methods can be found in Refs.~\cite{Nee2010.PRL104.160401,Nee2013.PRL111.235301}.    

\begin{figure}
\includegraphics[width=\linewidth]{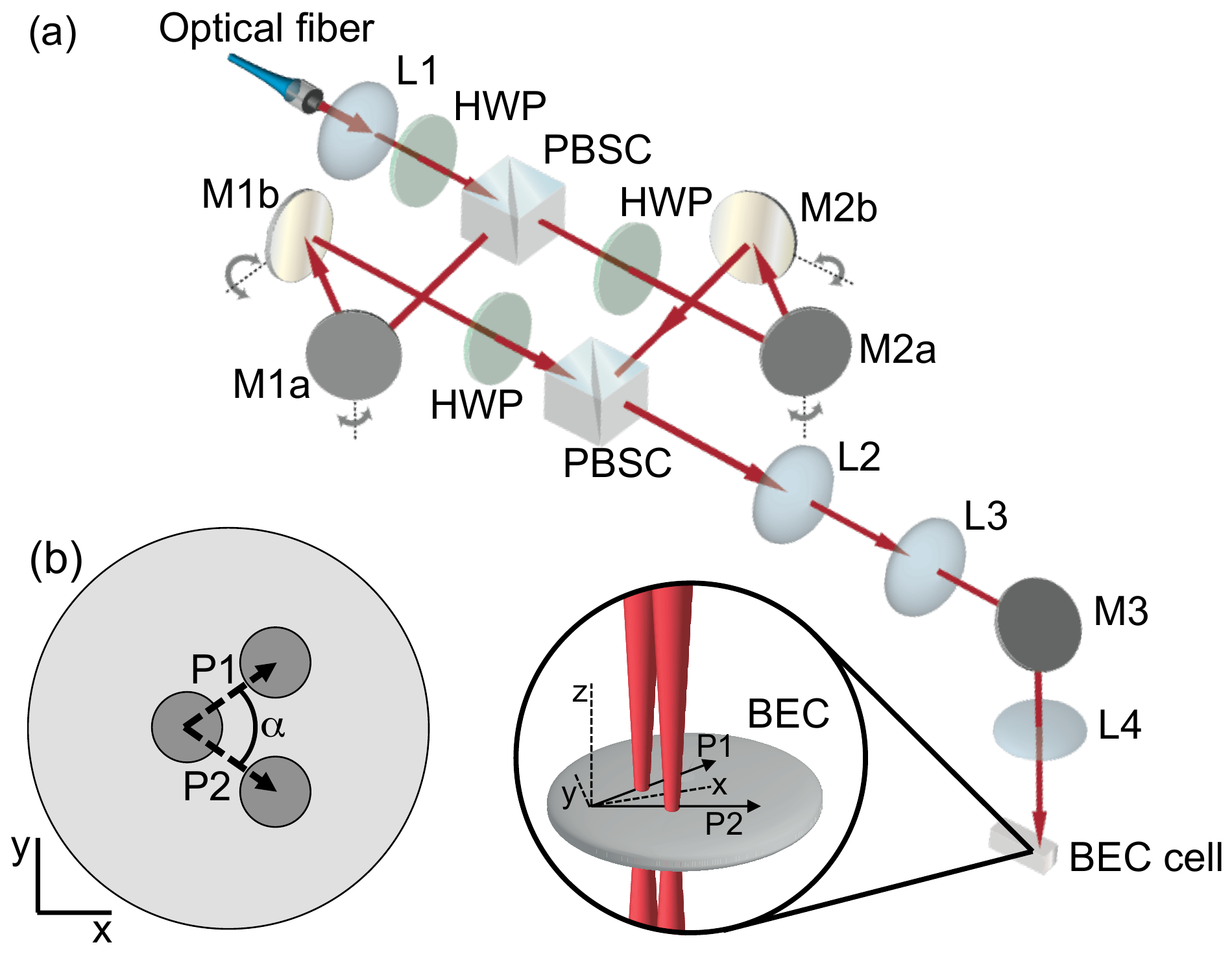}%
\caption{\label{}(Color online.)  (a) Diagram of the optical system that is used to translate two blue-detuned laser beams through the BEC. A 660-nm laser beam is brought into the optical system with a polarization-preserving single-mode optical fiber.  The beam is collimated by lens L1 and then split into two paths by a polarizing beamsplitter cube (PBSC); the optical powers of the two beams are balanced using a half-wave plate (HWP) placed before the first PBSC. Each beam reflects off of two mirrors (M1a and M1b, or M2a and M2b). Each mirror has a piezo-electric transducer (PZT) stack attached to an adjustment screw; the two PZTs attached to the mirrors of a given beam path control the tilts of the two mirrors about orthogonal axes, enabling the two laser beams to be remotely steered.   After passing through additional HWPs, allowing for further optical power adjustment, the two beams are recombined with another PBSC.   Lenses L2 and L3 comprise a 6$\times$  minification system in order to increase the angular deflection of the chopsticks beams, prior to deflection of the beams by mirror M3 and focusing the beams at the BEC by lens L4.  In the experiments reported here, the beams have $1/e^2$ radii at the BEC of $\sigma_c \sim19\,\mu$m.  During the course of an experimental run, the two laser beams move along paths $P1$ and $P2$ in the $(x,y)$ plane, as represented in the inset diagram.  (b) An illustration of the BEC (large gray disk) lying in the $(x,y)$ plane.  The laser beams are initially co-located and pierce the BEC (left dark-gray disk) and move along paths $P1$ and $P2$, which subtend an angle $\alpha\sim66^{\circ}$.  The beams end up in different locations (upper and lower right dark-gray disks).  The size of the laser beams, relative to the BEC Thomas-Fermi radius, and the initial and final locations of the laser beams, are illustrated to scale.}
\end{figure}

\begin{figure}
\includegraphics[width=\linewidth]{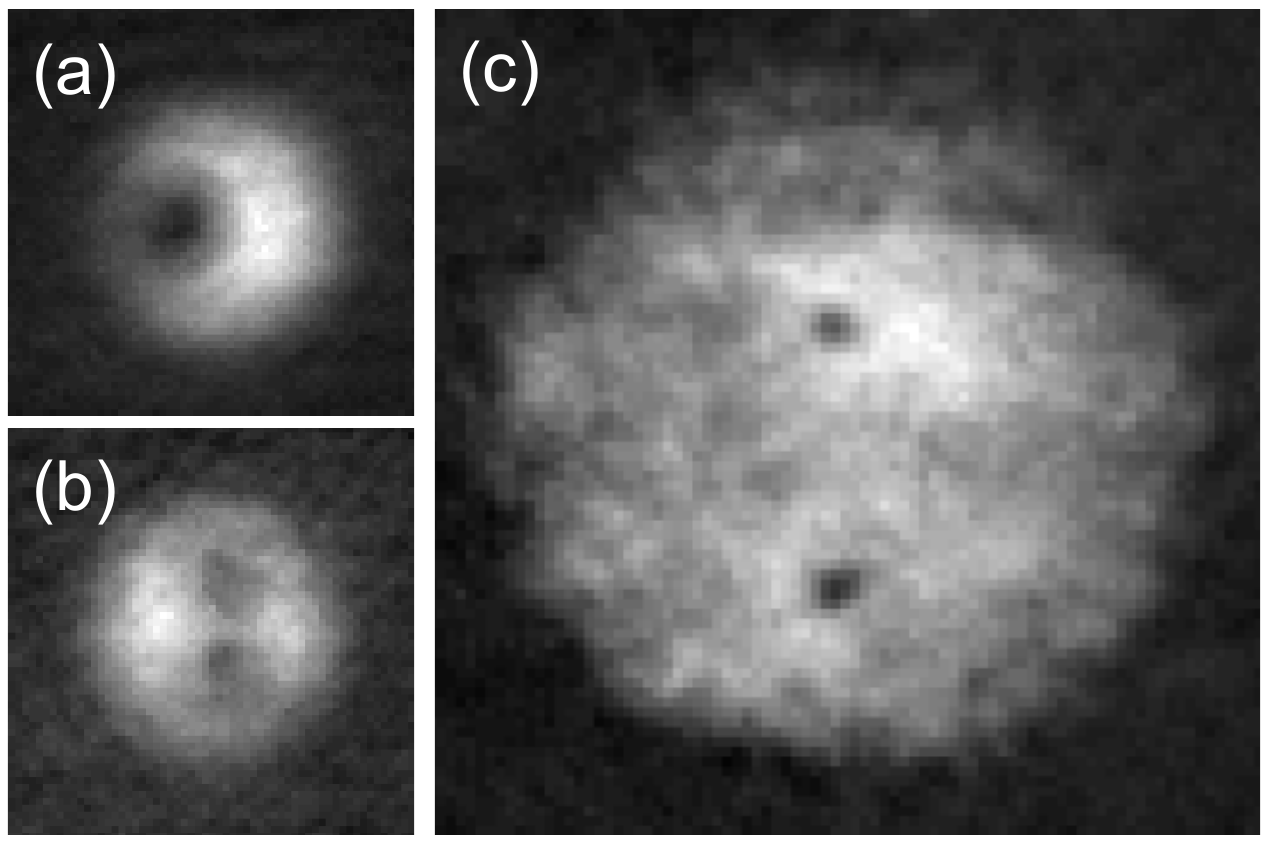}%
\caption{\label{}
 (a)  A 144-$\mu$m-square \emph{in situ} image of the BEC taken along the vertical ($z$) axis, showing the profile of the $\sim$100-$\mu$m-diameter BEC in the $(x,y)$ image plane.  The hole to the left of center is created by the two co-located chopsticks beams turned on at full power prior to moving them across the BEC.  (b) Same as (a), but with the chopsticks beams at their final locations at the ends of paths $P1$ and $P2$.  The intensity minima due to the chopsticks beams are $\sim31\,\mu$m apart in this image, slightly larger than the separation obtained when measuring the beam positions directly. (c) A 205-$\mu$m-square absorption image of a BEC at the end of the chopsticks procedure, after the chopsticks beams were ramped off at their final positions and the BEC was released from the trap and allowed to ballistically expand a factor of $\sim1.78$ in the radial direction.  Vortex cores appear as the two small dark holes in the density distribution, and are $\sim65\,\mu$m apart in this image, compared with an expected separation of $\sim55\,\mu$m given the final positions of the chopsticks beams and the expansion factor of the BEC.}
\end{figure}

The primary tool used in the chopsticks technique is a pair of blue-detuned focused laser beams propagating along the $z$ direction.  The beams serve as repulsive barriers that penetrate the BEC and whose positions can be translated within the plane of the BEC.   Figure 1 shows a schematic diagram of the optical system that is used to control the motion of these two beams; a detailed description of the optical system is provided in the figure caption.     The chopsticks beams are initially co-located at position $(x_i,\,y_i)=(-10,\,0)\,\mu$m relative to the center of the trap.  Prior to the final stage of evaporative cooling that creates a BEC, each beam is turned on to full power, corresponding to a peak repulsive potential of $\sim 0.8\,\mu_0$.   By creating the BEC after the beams are in place, we minimize excitations of the BEC that would otherwise result from turning on the beams after creating a BEC, and also damp out BEC sloshing in the harmonic trap that can otherwise decrease the effectiveness and repeatability of vortex creation.  Figure 2(a) shows an \emph{in situ} image of the BEC with the two chopsticks beams turned on at full power at their initial locations.  

To generate two vortices of opposite charge pinned to the chopsticks beams, we simultaneously move both beams along linear trajectories at a constant speed for a sweep time of $t_s = 0.7$ s.  Beam B1, in the upper half-plane, moves to a final position of $(x_f,\,y_f)=(13,\,15)\,\mu$m.  Beam B2, in the lower half-plane, moves to a final position $(x_f,\,y_f)=(13,\,-15)\,\mu$m.  An \emph{in situ} image of the BEC after the beams have been moved is given in Fig.~2(b).  The beams travel distances of $\sim27\,\mu$m at a speed of $\sim 34\,\mu$m/s, much less than the maximum speed of sound for our BECs of $c_0\sim1700\,\mu$m/s, and still much less than $\sim c_0/10$, the speed required to nucleate and shed a vortex dipole for comparable beam and BEC parameters \cite{Nee2010.PRL104.160401}.   Once the beams reach their final positions, their positions and powers are held constant for times up to 2.5 s, after which their optical powers are linearly decreased to zero in 0.25 s.  This rate at which the beams are linearly ramped off was determined experimentally, such that it is fast enough to ensure that the vortices would not move significantly far away from the beam positions as the beam power is decreased, but slow enough such that the BEC is not significantly excited by a rapid change in the potential well profile.  The BEC is then released from the trap and allowed to ballistically expand in order to look for the presence and positions of vortices using standard absorption imaging techniques.   

Fig.~2(c) shows a representative image in which two vortices were observed after applying the chopsticks procedure, with each vortex being located near the final position of one of the laser beams.  In one test of the repeatability of the chopsticks procedure, we created 30 BECs and subjected each to the procedure described above.  In 26 (86\%) of the cases we found two vortices, with each located at a position that closely matches the final beam position, relative to the BEC center.  In the remaining cases, either one vortex was observed in the BEC, or an extra unpinned vortex was observed in addition to the two vortices that were located at positions matching those of the chopsticks beams.  A likely explanation for these cases is the presence of spontaneously formed vortices \cite{Wei2008.Nat455.948}.  Such vortices can alter the background flow profile of the BEC such that the necessary conditions for vortex nucleation and pinning parameters are not achieved, or might annihilate with one of the vortices created by the chopsticks technique, or could remain present throughout the chopsticks mechanism of vortex creation and be visible as a third vortex.

\begin{figure}
\includegraphics[width=.94\linewidth]{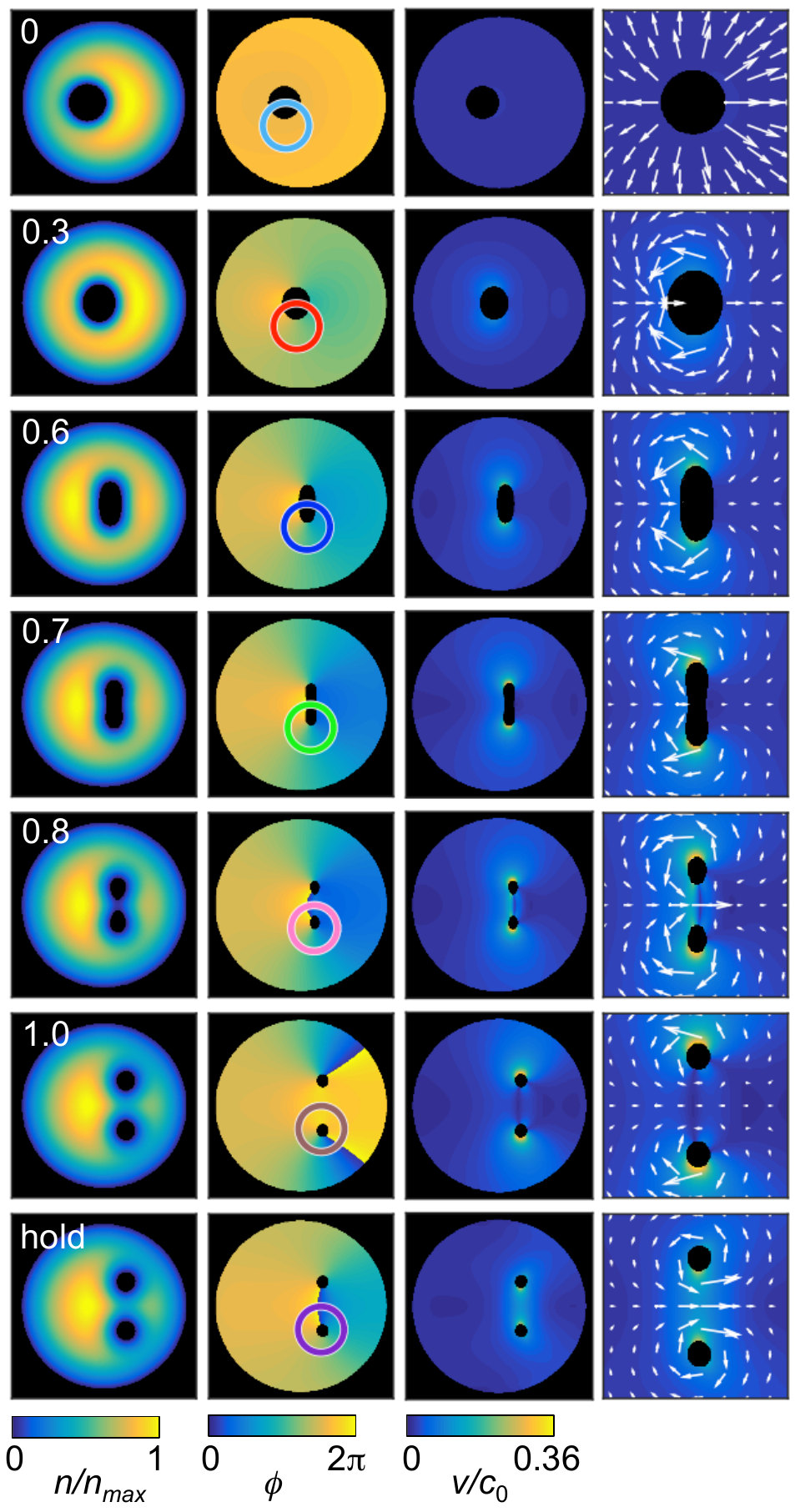}%
\caption{\label{}(Color online.) Results of 2D GPE numerical simulations of the chopsticks technique for simulation parameters described in the text.  Each row represents the state of the system at a time $t$ during the sweep, with the sweep time $t/t_s$ indicated in the upper left corner of the images of the left column.  The final row represents the state of the system at a hold time of 0.5 s after the end of the sweep, during which time the chopsticks beams' positions and powers remain unchanged from their final values.  Columns one through three show (respectively) for each time step 120-$\mu$m-square regions of the density profile $n$ relative to the maximum number density $n_\mathrm{max}$,  the quantum phase profile $\phi$, and the velocity profile relative to the maximum speed of sound $c_0 = \sqrt{\mu_0/m}$ for the ground state BEC without chopsticks beams present.  Column four shows a magnified view of the velocity profile, a 61-$\mu$m-square region centered on the point midway between the chopsticks beams.   Arrows indicate the local velocity of fluid flow at the tail of each arrow. For clarity, in all columns, points for which $n/n_\mathrm{max} < 0.0005$ are assigned the color black.  The colored circles in the second column are described in the caption of Fig.~4.  }
\end{figure}

\section{IV. Numerical Simulations}

To better understand the mechanisms at work in the chopsticks technique, we turn to numerical simulations of the Gross-Pitaevskii equation (GPE) \cite{Pet2008.Bose.book}.  We first assume a three-dimensional wavefunction of the form
\[ \Psi(x,y,z,t) = (\pi l_z^2)^{-1/4}\, \psi(x,y,t)\, e^{-z^2/2 l_z^2} \]
where the dynamics of the 2D macroscopic wavefunction $\psi(x,y,t)$ are assumed to closely approximate the dynamics of our actual three-dimensional (3D) system.  This approximation is appropriate for our highly oblate system, particularly because for our conditions, vortices remain aligned along the $z$ direction without undergoing significant tilting or bending \cite{Roo2011.PRA84.023637}.  The constant $l_z$ is an effective length scale for the axial thickness of our BEC, described below, and is not necessarily equal to the harmonic oscillator length for the $z$ direction.  

With this 2D approximation, the 3D GPE with a dimensionless phenomenological damping constant $\gamma$ can be reduced to 
\[ (i-\gamma)\frac{\partial}{\partial t} \psi = -\frac{\hbar^2}{2m} \nabla_{x,y}^2 \psi + V_{\mathrm{ht}} + V_\mathrm{c} + g_\mathrm{2D}|\psi|^2\psi,\]
where $V_{\mathrm{ht}} = \frac{1}{2}m{\tilde{\omega}}_r^2 (x^2 + y^2)$ is the potential energy of a 2D harmonic trap of frequency ${\tilde{\omega}}_r$, and $m$ is the mass of $^{87}$Rb. 
The potential due to the chopsticks beams is given by 
\[ V_\mathrm{c}(x,y,t) = U_0 \sum_{j=1}^2  \exp\{-2 [(x_i, y_i)_j + (v_x,\,v_y)_j t]^2/\sigma_\mathrm{c}^2\}   \]
where $U_0 = 0.8 \mu_0$ is the maximum repulsive potential energy of each beam during the sweep, $(x_i, y_i)_j$ is the initial position of beam $j$,  $(v_x,\,v_y)_j$ is the sweep velocity vector for beam $j$, and $\sigma_\mathrm{c} = 19\,\mu$m is the radius of each chopsticks beam.  In order to mimic the effects of non-zero temperature and the damping provided by a small thermal cloud of atoms, $\gamma$ is set to 0.003 for all simulations reported here \cite{Yan2014.PRA89.043613,Coc2011.PRA84.043640}.  

We use a 2D interaction parameter $g_\mathrm{2D} = \frac{4\pi\hbar^2 a}{m\sqrt{2\pi}l_z}$, where $a$ is the atomic $s$-wave scattering length for our atoms.  For the 2D GPE simulations, we use an effective length parameter $l_z = 0.53\,\mu$m, an effective radial trap frequency of ${\tilde{\omega}}_r=0.84\,\omega_r$, where $\omega_r/2\pi = 8$ Hz as in the experiment, and an effective number of atoms equal to $3.2\times 10^5$.  These numbers are chosen such that when the BEC is in its ground state without chopsticks beams present, the density at the center of the BEC in the 2D approximation matches that of a 3D Thomas-Fermi approximation for the actual experimental parameters, and the 2D Thomas-Fermi radius of $\psi(x,y)$ is also the same as that of the 3D Thomas-Fermi approximation.   This approach enables us to compare the beam velocities of the simulation to a speed of sound that is the same as calculated for our 3D experimental parameters in the Thomas-Fermi limit,  $c_0 = \sqrt{\mu_0/m}$, and permits direct comparison of beam radii and positions in the simulations with the equivalent parameters of the experiment.

Using a split-step routine, we simulate the dynamics of $\psi(x,y,t)$ on a 120-$\mu$m-square area using a 512~$\times$~512 grid, giving a grid point spacing of 0.23 $\mu$m that is comparable to the bulk healing length of $\sim 0.28\,\mu$m calculated for our experimental parameters.  After finding the ground state of the system in the presence of the static chopsticks beams at position $(x_i, y_i)_1 = (x_i, y_i)_2 = (-10,\,0)\,\mu$m, the beams are swept to their final positions $(13,\,\pm15)\,\mu$m in a sweep time $t_s=0.7$ s, as in the experiment.  Finally, with the beams remaining at their final positions and at their full powers, we propagate the simulation for an additional hold time of 0.5 s.  The results of this simulation are shown in Fig.~3.

We see from the simulation results of Fig.~3 that when the two beams start moving, fluid flow is induced in the BEC.  For example, at $t/t_s=0.3$, the velocity profile shows the highest fluid flow speeds at the sides of the moving barrier, corresponding to the regions of the largest phase gradients seen in the phase profile.   As the chopsticks beams eventually separate, between $t/t_s=0.7$ and $t/t_s=0.8$, a channel opens up between the beams, allowing the atomic superfluid in the regions in front of and behind the beams to merge. In order for $\psi$ to be single-valued and continuous in a path around each laser beam, the two merging regions of superfluid must adopt a continuous phase profile in the channel between the two beams.  Whether or not the path around a given laser beam encloses net circulation will depend upon how the regions merge and a phase gradient is established  \cite{Sch2007.PRL98.050404}.  As shown in Fig.~4, prior to beam separation and fluid merging, there is a phase difference of approximately 1.75$\pi$ established between the regions in the front and rear of beam travel; once the the beams separate and the front and trailing regions merge, the phase relaxes to an azimuthal phase winding of $2\pi$ ($-2\pi$) around beam B1 (B2), indicating the presence of a positively (negatively) charged vortex pinned by beam B1 (B2).  

\begin{figure}
\includegraphics[width=\linewidth]{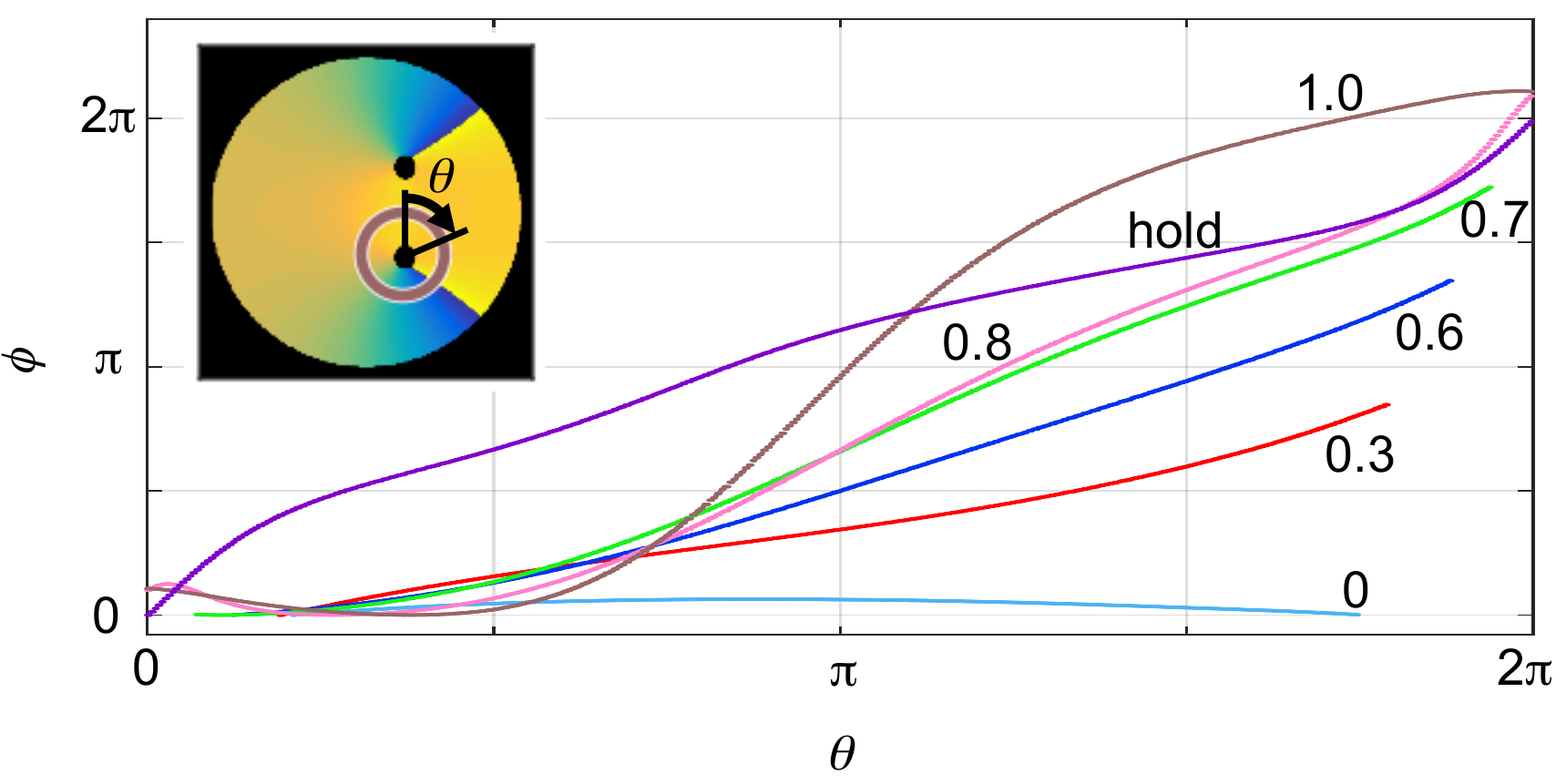}%
\caption{\label{} (Color online.)  For each time step $t/t_s$ represented in Fig.~3, the quantum phase $\phi$ around the path denoted by the colored circles in the phase profiles of Fig.~3 is shown here as a function of the angle $\theta$ around the circle.   The point $\theta=0$ is taken to be at the midpoint between the two laser beams, and $\theta$ increases in a clockwise direction, as indicated by the inset figure showing the phase profile at $t/t_r=1.0$.  The plots of $\phi$ vs.~$\theta$ in this figure are labeled with the values $t/t_s$ of the corresponding rows of Fig.~3, and the colors of the corresponding rings that indicate the points for which $\phi$ is shown.  At points where the density is negligible (i.e., for black-colored values in the phase profiles of Fig.~3), values for $\phi$ are not shown.  This plot shows the gradual establishment of a clockwise $2\pi$ phase winding around the lower laser beam.  At time $t/t_s=0.7$ (green line), immediately before the beams separate, a phase winding close to $2\pi$ has been established.  At $t/t_s=0.8$ (pink line), just after the beams separate, a continuous loop of fluid is seen to exist around each beam (see first column Fig.~3), with a phase winding of $2\pi$ around each beam.    For ease of comparison of the phase at different times $t/t_s$, each plot of $\phi$ vs.~$\theta$ shown here has a constant value added such that $\phi$ is continuous from $\theta=0$ to $\theta=2\pi$ and the minimum value of $\phi$ over this range is zero. }
\end{figure}

The phase winding around each beam at the end of the chopsticks process depends on the fluid flow speed induced by beam motion.  If, immediately prior to merging, a phase difference precisely equal to $\pi$ has gradually built up between these two regions (due to, for instance, slower beam velocities than used in our experiment), the system will support the temporary existence of a dark soliton between the two beams  \cite{And2008.ENPBEC.85}.   If the phase difference between the front and trailing regions is less than $\pi$, as is the case in Fig.~5 for $t_s = 1.4$ s, no vortices will be created or pinned.    If that phase difference is somewhat greater than $\pi$ and less than $3\pi$, the system will relax to one with a positively charged vortex pinned to beam B1, and a negatively charged vortex pinned to beam B2, as seen in rows 2 through 4 of Fig.~5.  Numerically, the presence of a small but non-zero damping constant $\gamma$ aids this relaxation process, but is not essential to the creation of vortices or their pinning at a laser beam.  For even larger phase differences, induced by faster sweeps, larger amounts of vorticity are created, but the vortices are not stably pinned by the laser beams for the conditions studied.  The numerical results shown in Fig.~5 demonstrate the robustness of the chopsticks method to substantial variations in the sweep time.

\begin{figure}
\includegraphics[width=.92\linewidth]{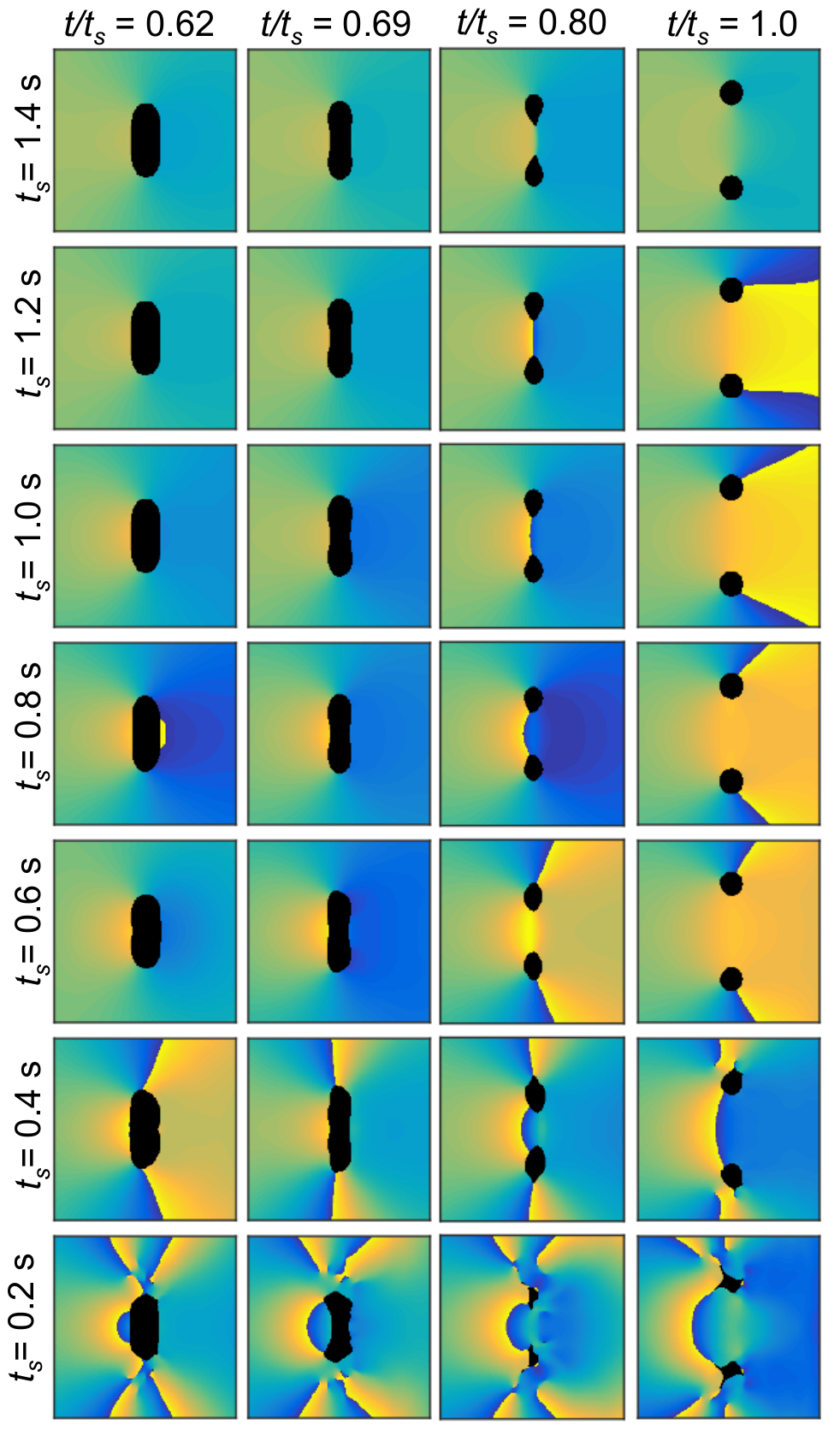}%
\caption{\label{} (Color online.)  Each column shows 61-$\mu$m-square regions of the phase profile $\phi(x,y)$ centered on the midpoint between the chopsticks beams for different times $t/t_s$, indicated above each column.  Sweep times $t_s$ are indicated at the left of each row.  The display colors are mapped to a range of 0 to $2\pi$, as in the second column of Fig.~3.   Except for the variation of $t_s$, simulation parameters are identical in all cases to those described in the text.  For $0.6 \leq t_s \leq 1.2$, two vortices are generated and left pinned to the beams, an indication of the robustness of the chopsticks technique.  For $t_s > 1.2$, an insufficent phase gradient is established around each beam, and vortices are not created.  For $t_s < 0.6$,  too large of a phase gradient is established, and two or more vortices are nucleated for each beam.  For the conditions studied, these additional vortices did not remain pinned to the laser beams, as can be seen from the singularities in the phase profiles in the last two rows. }
\end{figure}

\section{V. Manipulating Vortices}

As an experimental demonstration of deterministic manipulation of vortices, after generating and pinning two vortices and then holding the beam positions and powers fixed for 0.5 s, one of the beams was translated back to its original position over a return time of $t_r = 1.0$~s while the other beam remained stationary.   The beams were then held in these new positions for an additional time of 1.0 s.  The insets to Fig.~6 show \emph{in situ} images of the BEC with one of the two chopsticks beams returned  to its initial position after completion of the chopsticks procedure.  After ramping off the beams in 0.25 s, and releasing the BEC from the trap, two vortices were observed in the BEC at locations that correspond to the new final positions of the chopsticks beams, as shown in the main images of Fig.~6.

\begin{figure}
\includegraphics[width=\linewidth]{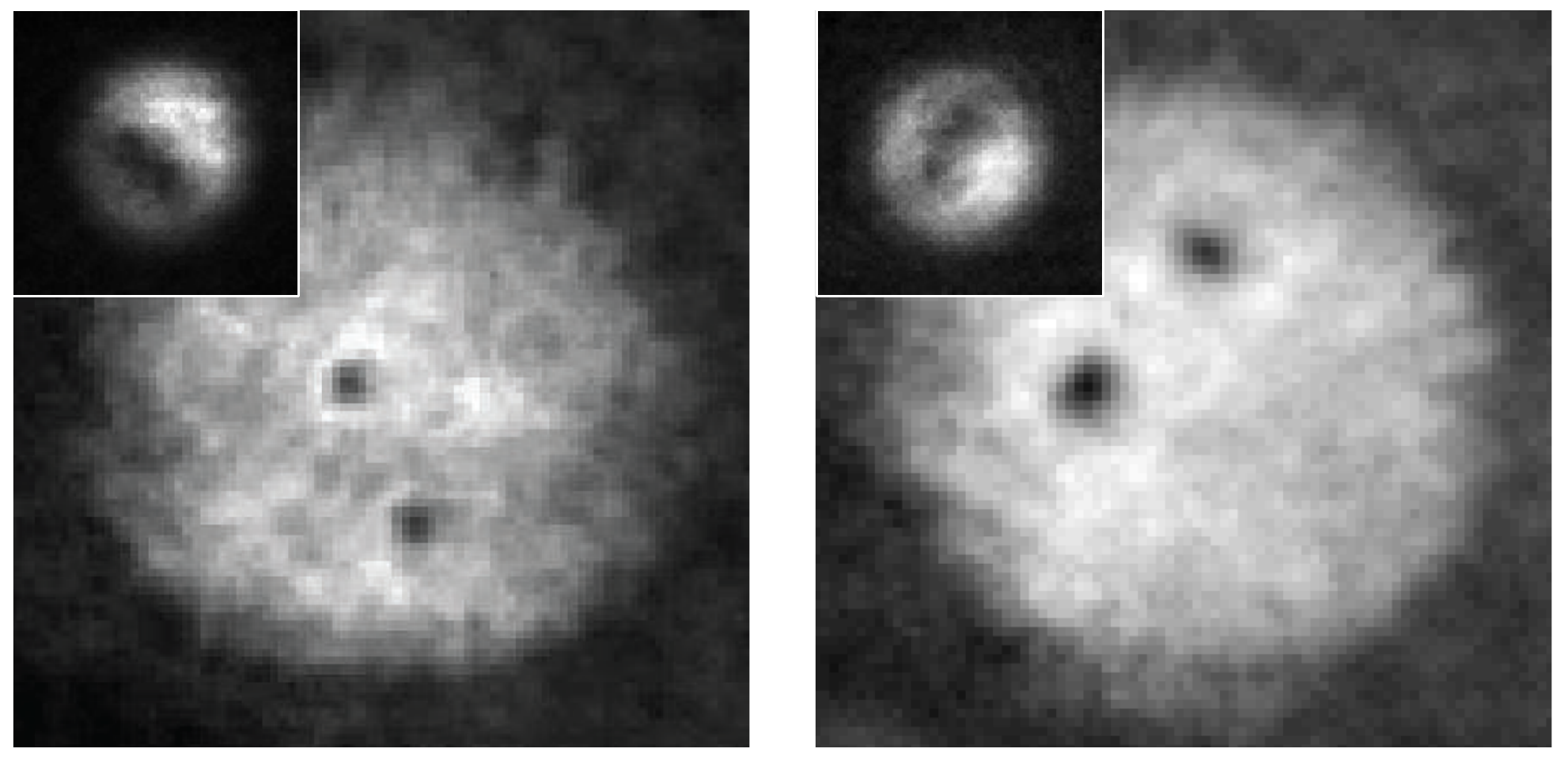}%
\caption{\label{} Experimental demonstration of vortex position manipulation.  After creating and pinning the two vortices, and waiting an additional 0.5 s, one of the chopsticks beams was returned to its initial position in 1.0 s while the other beam remained in place.  For the left image and inset, the upper beam (B1) was returned to its initial position, as shown in the inset \emph{in situ} image.   The inset for the right image shows the result of returning the lower beam (B2) to its initial position.   In each case, after a beam returned to its initial position, both beams were ramped off, releasing the vortices into the BEC.  The main images of this figure show the corresponding positions of the vortices after the BEC was released from its trap and imaged following a period of ballistic expansion.  }
\end{figure}

The chopsticks beams can also be recombined as a check that the equal and opposite circulations can be cancelled out on-demand.   We experimentally examined two different methods of returning the beams.  First we examined sequential beam motion, with one beam returning to the initial position $(x_i,\,y_i) = (-10,\,0)\,\mu$m in 1.0 s, followed by a 0.5 s hold during which both beams were stationary, followed by the second beam returning to the initial position in 1.0 s.  We also examined simultaneous beam motion, where both beams returned to the initial location in 1.0 s.   As expected, after returning \emph{both} chopsticks beams to their original position, either sequentially or simultaneously, vortices were repeatedly absent from the BEC. From our experimental data, however, it is not clear whether vortex-antivortex annihilation occured within the bulk BEC after vortices are pulled from the chopsticks beams, or if the vortices remained pinned to the lasers during the circulation cancellation process.   

To further examine dynamics of circulation cancellation, we performed a numerical simulation of the simultaneous return of the beams to the initial position in a return time of $t_r =1.0$ s.  From the simulation, shown in the top row of Fig.~7, we see that the vortices are first ripped away from the chopsticks beams within a time of $t/t_r = 0.11$.  Additional vortices are also seen around each beam's position, but none of these are pinned by the beams.  Eventually all vortices annihilate one another, and the system is left free of any fluid circulation.

We interpret this process as arising from the laser beams now moving against the background fluid flow initiated by the motion of the chopsticks beams, i.e. the fluid flow corresponding to the vortices now pinned to the beams.  At a nominal time $t/t_r = 0.1$ during the beam return, the beams are approximately $13.5\,\mu$m apart.  If vortices of opposite circulation are pinned to the beams, there is a background flow speed of $\sim 52\,\mu$m/s at each beam due to the presence of the vortex that is pinned to the opposite beam.  If the beams return to their initial position over distances of $27\,\mu$m in 1.0 s, they increase the speed of a beam relative to the background flow speed by nearly $27\,\mu$m/s.  The presence of a vortex pinned to a beam further increases the relative flow speed at the edge of each beam that is closest to the opposite beam (i.e, at the $\pm y$ boundaries of the channel between the beams).  Apparently, these conditions are sufficient to rip vortices off of the beams, and soon thereafter, to cause the nucleation of a vortex dipole at each beam, with all vortices then annihilating pairwise.

\begin{figure}
\includegraphics[width=\linewidth]{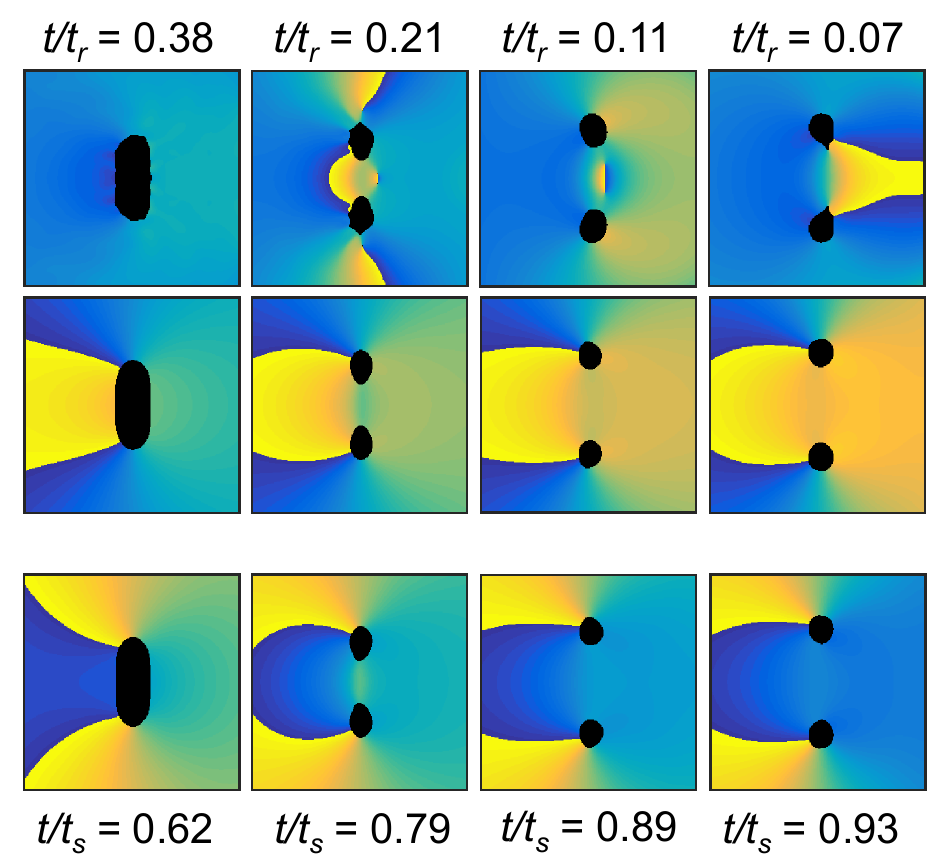}%
\caption{\label{} (Color online.)  Numerical results of 2D GPE simulations showing the return of both chopsticks beams to their initial positions over a return time $t_r$, for beam and BEC parameters the same as those of Fig.~3.  Phase profiles are shown for 61-$\mu$m-square regions centered on the midpoint between the chopsticks beams, with display conditions identical to those of the second column of Fig.~3.  For ease of comparison with the forward sweep simulations shown elsewhere in this paper, and specifically for the vortex generation procedure for a sweep time $t_s=0.7$ s, shown in the third row for sweep conditions identical to those of Fig.~3, the snapshots are given from left to right in reverse-time order.  The top row shows the return of the beams after a 0.5 s hold time, with $t_r=1.0$ s.  Beams are moving from right to left, and are in the process of merging with one another.  At time $t/t_r=0.07$, immediately after the beam return starts (top row, rightmost image), the vortices are still co-located with each laser beam.  Soon afterwards, at  $t/t_r=0.11$, the vortices are clearly no longer pinned (top row, third column).  Additional vortices are subsequently nucleated (top row, second column) and annihilated, leaving a state with no circulation by $t/t_r=0.38$ (top row, first column).  For the same return sweep, but with a conjugation of the quantum phase (second row, also see text), the vortices remain pinned to the beams until the beams recombine.  This return process is nearly identical (aside from the phase conjugation) to the initial vortex generation stage, shown in the third row for beam separations identical to those of the first two rows, but presented for increasing times from left to right.}
\end{figure}

In a second numerical study of the return of both beams to the initial locations, we again performed the simulation described above, but first conjugated the BEC phase at the end of the chopsticks procedure.  Phase conjugation is equivalent to reversing the charge of each vortex, or to slowly exchanging the positions of the two beams by rotating them around their center point, which was also examined numerically and is experimentally feasible.  After phase conjugation, the beams were returned to their initial positions in 1.0 s.   Because the beams were now moving in roughly the same direction as the background flow, the speed of the beams relative to the background flow was much less than   $\sim 52\,\mu$m/s and the vortices remained trapped until the opposite circulations canceled each other when the beams recombined, as shown in the second row of Fig.~7.   Furthermore, as the third row of Fig.~7 shows, the phase profiles for the  times at which the chopsticks beams have the equivalent spatial separation during the initial forward chopsticks sweep are remarkably similar to the return sweep after the phase conjugation step.  We therefore conclude that while the chopsticks procedure cannot be simply reversed in order to return the system to the initial state, the process does demonstrate  reversibility when vortex charges are also first reversed.  This charge-time symmetry is similar to the hysteresis recently reported when spinning up or slowing down the persistent superfluid current in a toroidally trapped BEC \cite{Eck2014.Nat506.200}.

\section{VI. Conclusions and Outlook}

The primary results of this paper are the proof-of-principle experimental and numerical demonstrations of a deterministic method of creating and manipulating, on-demand, two vortices of opposite circulation in a BEC.  There are numerous parameters involved in determining the range of utility of this method, and here we have begun to explore some of those parameters.  Our demonstrations are a step towards the engineering of arbitrary vortex distributions in a BEC, and also open up a new pathway to precision studies of the interactions of vortices with each other, with sound, with potential barriers and obstacles in a BEC, and as a function of other parameters such as trap geometry and temperature.   To more fully understand how this technique can be best utilized for studies of vortex dynamics and vortex interactions, additional experimental and numerical research is required, particularly regarding an assessment of the motion of the vortices once the chopsticks beams are ramped off, and the ranges of beam powers, sizes, and sweep speed that can be successfully used in the chopsticks process.  

The goal of creating specified distributions of many vortices, on-demand, remains an experimental challenge, but one that is within reach.  From numerical explorations,  it appears possible to simultaneously utilize multiple pairs of chopsticks beams, or to use just two chopsticks beams with stationary pinning sites in successive vortex creation stages.   Additional beam manipulation procedures that remove vortices from a BEC can also be explored in order to create charge-imbalanced populations of vortices.  These aspects of the chopsticks technique are numerically explored and presented in a companion article \cite{Ger2015.ARXIV15xx.xxxx}.   

\section{Acknowledgments}

\begin{acknowledgments}

This research was supported by grants PHY-0855677 and PHY-1205713 from the U.S.~National Science Foundation.  K.E.W. acknowledges support from the Department of Energy Office of Science Graduate Fellowship
Program, administered by ORISE-ORAU under Contract No.~DE-AC05-06OR23100. Z.L.N. acknowledges partial support from the University of Arizona TRIF Program.  We thank R.~Carretero-Gonz{\'a}lez and P.G.~Kevrekidis for a critical reading of the manuscript.

\end{acknowledgments}


%

\end{document}